\documentclass[reprint,twocolumn,superscriptaddress,secnumarabic,amssymb, nobibnotes, aps, pra]{revtex4-1}

\usepackage{amsmath}    
\usepackage{graphicx}    
\usepackage{verbatim}    
\usepackage{color}           
\usepackage{subfigure}   
\usepackage{hyperref}    
\usepackage{verbatim}  
\begin{document}

\title{LDA+DMFT approach to core-level spectroscopy:  application to 3$d$ transition metal compounds}

\author{Atsushi Hariki}
\affiliation{Institute for Solid State Physics, TU Wien, 1040 Vienna, Austria}
\author{Takayuki Uozumi}
\affiliation{Department of Mathematical Sciences, Graduate School of Engineering,
Osaka Prefecture University 1-1 Gakuen-cho, Nakaku, Sakai, Osaka 599-8531, Japan}
\author{Jan Kune\v{s} }
\affiliation{Institute for Solid State Physics, TU Wien, 1040 Vienna, Austria}

\date{\today}

\begin{abstract}
We present a computational study of 2$p$ core-level X-ray photoemission spectra of
transition metal monoxides MO (M=Ni, Co, Mn) and sesquioxides M$_2$O$_3$ (M=V, Cr, Fe) 
using a theoretical framework based on the local-density approximation (LDA) $+$ dynamical mean-field theory (DMFT).
We find a very good description of the fine spectral features, which improves considerably over 
the conventional cluster model. We analyze the role of the non-local screening and its
relationship to the long-range magnetic order and the lattice geometry.
Our results reveal the potential of the present method for the analysis and
interpretation of the modern high-energy-resolution experiments.

\end{abstract}

\maketitle

\section{ Introduction}

Materials with strongly correlated electrons host a number of fascinating phenomena
ranging from the high-temperature superconductivity
to exotic orders of spin, orbit and charge degrees of freedom.
Microscopic understanding of the complex interplay between the formation of
atomic multiplets and inter-atomic hybridization --chemical bonding--
is one of the challenging topics in condensed matter physics \cite{imada98,khomskii14}. 
Core-level X-ray spectroscopy is a powerful tool 
for investigation of the strongly correlated materials \cite{groot_kotani}.
The last decade  witnessed a great advance of high-resolution and bulk-sensitive techniques for the first-order optical processes, 
such as X-ray photoemission spectroscopy (XPS) with hard X-ray ($>$5 keV),
\cite{Taguchi16,taguchi08,eguchi08,obara10,taguchi10,horiba04,kamakura04,miedema15}
as well as 
resonant inelastic X-ray scattering (RIXS) \cite{ament11,ghiringhelli09}.
The experimental progress opened access
to fine spectral features reflecting the low-energy physics,
e.g., elementary magnetic excitations
\cite{guarise10,kim12,minola15}.

Theoretical modeling is a crucial step in inferring the microscopic physics from experimental spectra. 
With XPS, the system is probed through a response to the core hole created by an X-ray irradiation.
The core hole, e.g., in the 2$p$ shell of a transition metal (TM), strongly interacts with localized 3$d$ electrons,
which leaves a fingerprint of the atomic multiplet structure in the spectra. 
In addition, the core hole presents a charge perturbation
which induces a dynamical response of the valence electrons -- charge transfer (CT) screening.
The CT screening effectively amplifies the effect of hybridization of the excited atom with its surroundings in the core-level XPS. 

MO$_6$ cluster model (CM) is probably the most popular model
conventionally employed to analyze X-ray spectra of TM compounds since 1980's \cite{zaanen86,okada91,bocquet96}.
In this model the intra-atomic interactions on the TM site 
and hybridization of the TM 3$d$ states with the neighboring ligands are considered,
while the rest of the lattice consisting of the ligand and TM atoms is neglected.
The CM has been very successful in explaining the overall structure of the XPS
and X-ray absorption spectra of numerous TM compounds. However, its limitations when it comes to the
fine spectral features became obvious with the arrival of high-resolution experiments.
For example, it fails to reproduce the fine structure
of the 2$p_{3/2}$ main line (ML) observed in a series of TMOs (transition metal oxide)
\cite{Taguchi16,taguchi08,eguchi08,obara10,taguchi10,horiba04,kamakura04,miedema15}.
A similar failure of the CM was reported for other excitation processes,
such as $L$-edge RIXS in TMOs \cite{agui09}.
It was proposed that the failure results from the absence of so-called nonlocal screening (NLS) \cite{veenendaal93}. 
The NLS involves the many-body states, which include the TM neighbors, responsible for the low-energy physics of spin and orbital ordering~\cite{veenendaal06,hariki13b,hariki16}.
Thus, it allows the core-level XPS to probe also non-local phenomena.

For the theory to keep up with the high-resolution experiments it is important to introduce
a framework, which overcomes the limitations of the CM analysis.
In this article, we present a systematic study of 2$p$ XPS spectra of selected 3$d$ compounds
based on the local-density approximation (LDA) $+$ dynamical mean-field theory (DMFT)~\cite{metzner89,georges96}.
The present approach \cite{hariki13a,hariki13b} consists in post-processing of the LDA+DMFT calculations,
in which the Anderson impurity model (AIM) with the DMFT hybridization functions
is extended to include explicitly the core orbitals 
and their interaction with TM $3d$ orbitals. 
Technically, the discrete ligand states of CM are replaced by a continuous
DMFT bath, which contains the information about the entire lattice.
Besides the conceptual advance, the method eliminates the ambiguities
in the choice of the CM parameters, which are replaced by (almost) parameter-free
LDA+DMFT calculation~\cite{kotliar06,kunes09}.
To calculate the spectra of the extended AIM 
an impurity solver based on the configuration-interaction scheme was developed~\cite{hariki15,hariki16}.

Previously, 
some of us applied the described approach to 
the 2$p_{3/2}$ XPS in cuprates, NiO and La$_{1-x}$Sr$_x$MnO$_{3}$ \cite{hariki13a,hariki13b,hariki16}.
A close relationship of the experimental features of the 2$p_{3/2}$ peak 
to the many-body composite structure of the top of the valance band, 
such as the Zhang-Rice band,
and long-range spin/orbit order was pointed out.
Here,
we report a systematic analysis of 2$p$ XPS spectra of MO (M=Ni, Co, Mn)
and M$_2$O$_3$ systems (M=V, Cr, Fe), with special
attention to the NLS. 
We discuss how the fine spectral features are related to the material specific properties,
such as the metallicity of V$_2$O$_3$, the magnetic order of NiO, CoO and Fe$_2$O$_3$
or to the crystal geometry,
on which the NLS is shown to depend sensitively.
Our results show that NLS is a common contributing factor to the core-level XPS spectra of TMO
and it must be taken into account when interpreting the spectra.


\section{Theoretical method}
\label{sec:2}
The calculation of the core-level spectra proceeds in three steps:
(i) construction of a $dp$ model from the converged LDA calculation,
(ii) solution of the DMFT self-consistent equation for the $dp$ model to obtain the DMFT hybridization function, and 
(iii) calculation of the core-level spectra using the extended AIM with the DMFT hybridization function.  
The steps (i) and (ii) are standard for the LDA+DMFT method.
In DMFT the local correlations are included explicitly, while the non-local
correlations are included only on the static mean-field level~\cite{georges96}.
The core of the method is mapping of the lattice problem onto
an AIM with self-consistently determined hybridization function $V(\varepsilon)$~\cite{georges92}.
The orbital- and spin-diagonal~\footnote{In a general case $V^2(\varepsilon)$ is a matrix.} hybridization density $V(\varepsilon)$ on real-energy axis is given by \cite{georges96,hariki13b},
\begin{equation}
   V_{\gamma \sigma}^2(\varepsilon)=-\frac{1}{\pi}{\rm Im}
        \langle d_{\gamma\sigma}|\bigl (\varepsilon-h^{0}-
        \Sigma(\varepsilon)-G^{-1}(\varepsilon)\bigr )
|d_{\gamma\sigma}\rangle,
\end{equation}
where $\Sigma(\varepsilon)$, $G(\varepsilon)$ and $h^0$ are 
the local self-energy, the local Green's function and the one-body part of the on-site Hamiltonian, respectively.
The $\gamma$ and $\sigma$ denote orbital and spin indices.

In step (i), 
we perform an LDA calculation with the WIEN2K package \cite{wien2k} for the experimental lattice parameters. 
The LDA bands of the TM 3$d$ and O 2$p$ are mapped onto a $dp$ tight-binding (TB) model
using the WIEN2WANNIER interference \cite{wien2wannier} and the WANNIER90 code \cite{wannier90}.
The spin-orbit (SO) interaction within the $3d$ shell turned out to have a negligible effect
on the studied spectra and the presented results were obtained without it.
When necessary, the inclusion of the SO interaction into the $dp$ model is straightforward \cite{wien2wannier}.

In step (ii), the TB model is augmented with the two-particle Coulomb interaction within
the TM 3$d$-shell and DMFT is employed to iteratively calculate the local self-energy and
the hybridization function. Merging LDA with many-body approached suffers from the 
well known problem how to avoid double counting the interaction terms. In this work,
we renormalize the $3d$ site energies by a constant shift $\mu_{d}$ treated as an adjustable 
parameter. The values of $\mu_{d}$ chosen such that the DMFT spectra reproduce well 
the valence band photoemission experiments are listed in Table~\ref{table:param}.
A long-range order, e.g., antiferromagnetic (AF) spin order, may develop
if the spin dependence of the self-energy and the proper magnetic unit cell are allowed.
We use the continuous-time quantum Monte Carlo method (CT-QMC) in the hybridization expansion algorithm~\cite{werner06,gull11} at this step.
The CT-QMC calculation is performed using a standard code \cite{hariki15,hariki16} based
on the segment picture with recent improved estimator techniques \cite{boehnke11,hafermann12}, with
the density-density form of the Coulomb interaction used for computational efficiency.
The Coulomb interaction between 3$d$ electrons is parameterized by $U=F_0$ and $J=(F_2+F_4)/14$,
where $F_0$, $F_2$ and $F_4$ are the Slater integrals \cite{eva11,krapek12}.
The configuration-averaged Coulomb interaction $U_{dd}$
is given as $U_{dd}=U-4J/9$.
Once the self-consistency is achieved,
self-energy $\Sigma(\varepsilon)$ on the real frequency axis is computed
by analytic continuation using the maximum entropy method \cite{jarrell96,wang09}.

Next, we construct the extended AIM. The hybridization density for real frequencies 
is obtained from (1) and approximated
by 30 bath states, which provides a reasonable consistency with the CT-QMC data.
The AIM is augmented with the
$2p$ core states. The $2p-3d$ interaction is parametrized 
with the Slater integrals.
These and the SO coupling within the $2p$ shell 
are calculated with an atomic Hartree-Fock code and the values are scaled down to 75\%$\sim$80\%
of their actual values to simulate the effect of intra-atomic configuration interaction from higher basis configurations \cite{matsubara05}.
Full Coulomb $3d-3d$ and $2p-3d$ interaction without any approximations is considered at this step.

{\renewcommand\arraystretch{1.4}
\begin{table}[t]
\begin{center}
  \begin{tabular}{l c  r  r r r r r r r |} \hline
                       &      NiO    & CoO  & MnO   & V$_2$O$_3$ & Fe$_2$O$_3$ & Cr$_2$O$_3$ & LaCrO$_3$ & \\ \hline
    $U$            &      7.0     &  7.3    &   7.0    &  4.8  &  6.8    &  6.4  & 7.0   \\
    $J$             &      1.1     &  1.1    &  0.95   &  0.7  &  0.86  &   0.8  & 0.8  \\  
    $U_{dc}$    &      7.8     & 8.6     &  8.5     &  6.5  & 8.4     &  9.0 & 9.0    \\
    $\mu_{d}$ &     52.0    &  47.6  &  30.5  &  8.1   & 30.6  &  21.3 & 23.8    \\  \hline
  \end{tabular}
  \vspace{-0.0cm}
  \caption{ Coulomb interaction $U$, Hund's interaction $J$, core-hole potential $U_{dc}$ and double counting correction $\mu_{d}$
  used in the studied compounds (in eV). }
  \end{center}
  \label{table:param}
    \vspace{-0.2cm}
\end{table}

The $2p$ XPS spectral function for the binding energy $E_B$ is given by
%
%
\begin{equation}
      F_{\rm XPS}(E_{\rm B})=
      -\frac{1}{\pi} {\rm Im} \sum_{n} \langle n | \hat{c}^{\dag}\frac{1}{E_{\rm B}+E_n-\hat{H}} \hat{c} | n \rangle
      \frac{e^{-E_{n}/k_{B}T}}{Z},
      \vspace{+0.3cm}
\end{equation}
where $E_{n}$ is the eigenenergy of $n$-th excited states $| n \rangle$ and $e^{-E_{n}/k_{B}T}/Z$
is the corresponding Boltzmann factor with the partition function $Z$.
The operator $\hat{c}$ creates a 2$p$ core hole at the impurity TM site.
The spectral function is calculated using the Lanczos algorithm within a configuration interaction scheme \cite{hariki15}.

The impurity Hamiltonian $\hat{H}$ has the form
\begin{equation}
  \hat{H} =  \hat{H}_{\rm TM} + \hat{H}_{\rm hyb},
\end{equation}
where $\hat{H}_{\rm hyb}$ describes hybridization with the fermionic bath \cite{matsubara05}.
The on-site Hamiltonian $\hat{H}_{\rm TM}$ is given as,
\begin{align}
\hat{H}_{\rm TM}
&=\sum_{\gamma,\sigma}\tilde{\varepsilon}_{d} (\gamma) \hat{d}_{\gamma\sigma}^{\, \dagger} \hat{d}_{\gamma\sigma} 
 + U_{dd} \sum_{\gamma\sigma > \gamma'\sigma'}
  \hat{d}_{\gamma\sigma}^{\, \dagger} \hat{d}_{\gamma\sigma} \hat{d}_{\gamma'\sigma'}^{\, \dagger} \hat{d}_{\gamma'\sigma'} \notag \\
& - U_{dc} \sum_{\gamma,\sigma , \,\zeta,\eta} 
  \hat{d}_{\gamma\sigma}^{\, \dagger} \hat{d}_{\gamma\sigma} (1-\hat{c}_{\zeta\eta}^{\, \dagger} \hat{c}_{\zeta\eta}) 
 +\hat{H}_{\rm multiplet}.
\label{eq:Imp_Hamiltonian}
\end{align}
Here, 
${\hat{d}_{\gamma\sigma}^{\, \dagger}}$ (${\hat{d}_{\gamma\sigma}}$) and ${\hat{c}_{\zeta\eta}^{\, \dagger}}$ (${\hat{c}_{\zeta\eta}}$) are
the electron creation (annihilation) operators  for TM 3$d$ and 2$p$ electrons, respectively.
The ${\gamma}$ (${\zeta}$) and ${\sigma}$ (${\eta}$) are the TM ${3d}$ (2$p$) orbital and the spin indices.
The TM $3d$ site energies $\tilde\varepsilon_{d}(\gamma)=\varepsilon_{d}(\gamma)-\mu_d$ are the energies
of the Wannier states $\varepsilon_{d}(\gamma)$ shifted by the double-counting correction $\mu_d$. 
The isotropic part the $3d-3d$ ($U_{dd}$) and $2p-3d$ ($U_{dc}$) interactions are shown explicitly,
while terms containing higher Slater integrals and the SO interaction are contained in $\hat{H}_{\rm multiplet}$.



\section{Results and  Discussion}
\label{sec:3}
The calculations were performed for temperatures of 300~K except for the PM phase in NiO at 800~K.

\subsection{NiO}

Figs.~\ref{fig:NiO}a,b show the Ni 2$p$ XPS calculated for the PM and AF phases,
respectively.
The large SO interaction in the $2p$ shell splits the spectra into well separated $2p_{1/2}$ and $2p_{3/2}$ parts.
Each of these is distinguished into two peaks transitionally called the main line (ML) at lower binding energy and
the charge-transfer (CT) satellite at higher binding energy.
These peaks exhibit an internal fine structure, most prominent of which is the 
double-peak $2p_{3/2}$ ML in the AF phase. Unlike the present approach,
the CM yields a sharp single-peak $2p_{3/2}$ ML~\cite{veenendaal93,bocquet92}.

Previously, some of us showed~\cite{hariki13b} that non-local screening (NLS) in a simplified
$d_{e_g}p$ model can account for the double-peak structure. Here, we extend this result
to the full $3d$-shell and provide a more detailed discussion of the effect. Numerical
experimenting with the hybridization function reveals that the low-$E_B$ peak
originates from NLS from the Zhang-Rice band, while the high-$E_B$ peak is a result
of local screening from neighboring O 2$p$ states  \cite{taguchi08,hariki13b}.
The corresponding final states of the XPS process may be denoted
$|\underline{c}d^{9}\underline{D}^{1}\rangle$ and $|\underline{c}d^{9}\underline{L}^{1}\rangle$,
where $\underline{c}$ , $\underline{L}$ and $\underline{D}$
represent a hole in the Ni 2$p$ core, in the O $2p$ band and in the Zhang-Rice band, respectively.
%
\begin{figure}
\begin{center}
   \includegraphics[width=80mm]{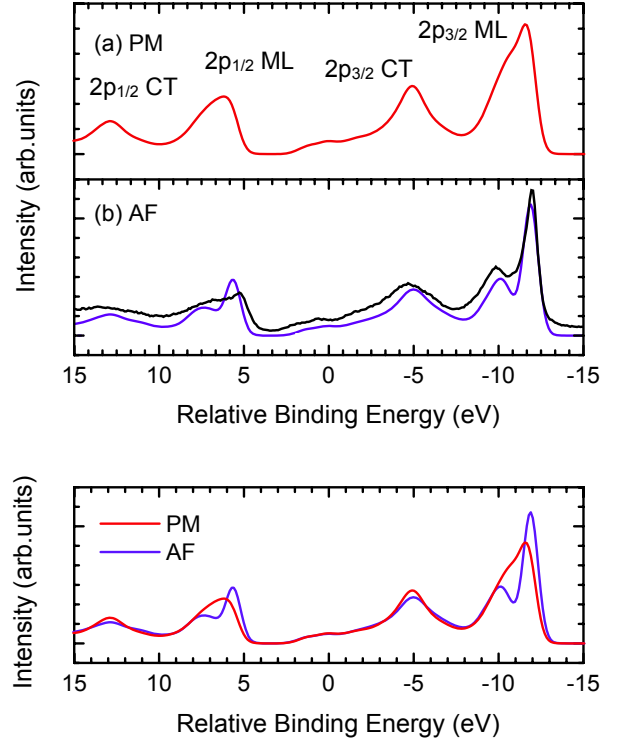}
\end{center}
\vspace{-0.5cm}
\caption{(Color online) 
The Ni 2$p$ XPS of NiO calculated for (a) PM  and (b) AF phase. 
The experimental data in~(b) is taken from Ref.~\cite{taguchi08}.
The spectra of the AF and PM phases are shown together (bottom), for comparison.
The spectral broadening is considered using a Gaussian of 0.5 eV width (HWHM).
}
\vspace{-0.4cm}
\label{fig:NiO}
\end{figure}
%
%
%
The $|\underline{c}d^{9}\underline{L}^{1}\rangle$ and $|\underline{c}d^{9}\underline{D}^{1}\rangle$ states 
in the many-body Hamiltonian repel each other due to the virtual hopping via the $|\underline{c}d^{8}\rangle$ state.
The splitting of the ML is more pronounced in the AF phase where
the NLS is enhanced relative to the PM phase, as investigated in Ref.~\cite{hariki13b}.
It is worth noting that the present approach also improves the description of the CT-satellite over the CM result.
This is because over-screened final states, such as 
$|\underline{c}d^{10}\underline{L}^{1}\underline{D}^{1}\rangle$ state,
overlap with the CT-satellite.
\begin{figure}

\vspace{+0.5cm}
\begin{center}
   \includegraphics[width=75mm]{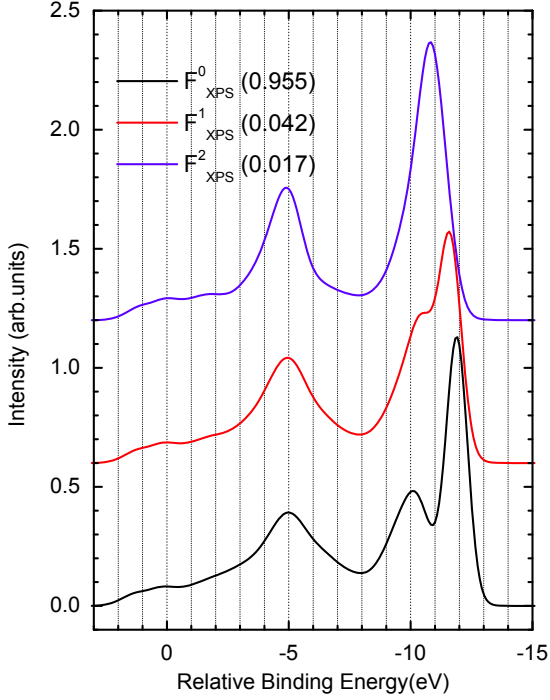}
\end{center}
\vspace{-0.5cm}
\caption{(Color online)
The Ni 2$p_{3/2}$ spectra in the AF phase of NiO
before the thermal average is summarized with the corresponding Boltzmann factors.
The spectral broadening is considered using a Gaussian of 0.5 eV width (HWHM).
}
\vspace{-0.0cm}
\label{fig:NiO_temp}

\end{figure}

To get more insight into the NLS mechanism, 
Fig.~\ref{fig:NiO_temp} shows the contributions $F^{n}$ to the AF 2$p_{3/2}$ XPS (before
multiplication with Boltzmann factors) from the three lowest-energy states, 
which are the exchange split members of the $S$=1 triplet~\footnote{Note that the $S$ and $S_z$ are not exact conserved quantities due 
to the SO interaction, but are still suitable to characterize the system.}.
While the splitting of ML is very distinct in the ground state contribution $F^{0}$,
the two peaks get closer to each other in $F^{1}$.
In $F^{2}$ with $S_z$=$-1$ character of $|n\rangle$ the ML becomes a single peak because 
NSL from the polarized Zhang-Rice states is forbidden by Pauli principle.
In NiO at 300~K the effect of thermal averaging is minor and the spectrum is dominated by the ground state.
That this is not always the case even in insulators is shown by our next example, CoO.


\subsection{CoO}

\begin{figure}

\begin{center}
   \includegraphics[width=75mm]{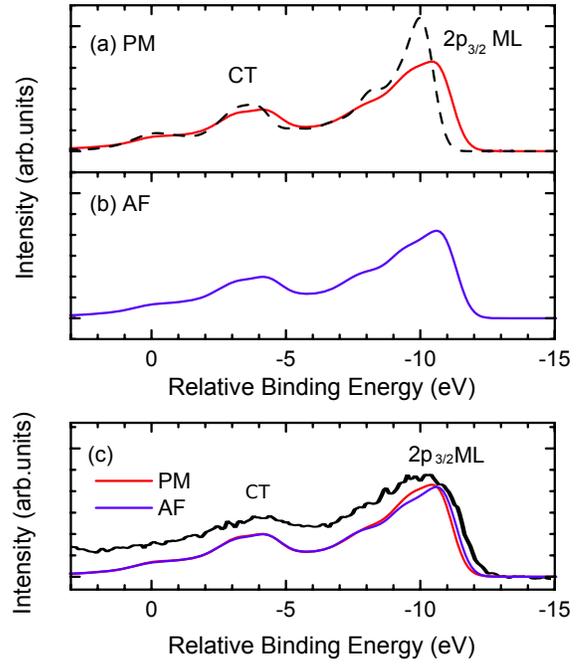}
\end{center}
\vspace{-0.5cm}
\caption{(Color online)
The Co 2$p_{3/2}$ XPS calculated for (a) PM and (b) AF phase at 300 K.
The spectrum obtained by CM is shown by dashed curve in (a).
The spectra in the PM and AF phases are shown together in (c), for comparison.
The spectral broadening using a Gaussian of 0.5 eV width (HWHM) is considered.
The experimental data in~(c) is taken from Ref.~\cite{chainani04}.
}
\vspace{+0.2cm}
\label{fig:CoO}

\end{figure}

In Fig.~\ref{fig:CoO}, we compare Co 2$p_{3/2}$ XPS in CoO
for (a) PM and (b) AF phase obtained in the present approach, with the experimental
spectra \cite{chainani04} and with the CM calculation. 
The NLS from the states at the valence band top, 
absent in the CM description, leads to broadening of the ML,
but unlike in NiO does not produce any distinct peaks.
Indeed, hard X-ray measurements found an anomalously
broad 2$p_{3/2}$ ML  with the shoulder in spite of high-energy resolution \cite{chainani04}.
Comparing the spectral in the AF and PM phases in Fig.~\ref{fig:CoO}c,
we find only a minor dependence on the magnetic order, which is
consistent with the measurements across $T_{\rm N}$ \cite{shen90}.

\begin{figure}
\begin{center}
   \includegraphics[width=75mm]{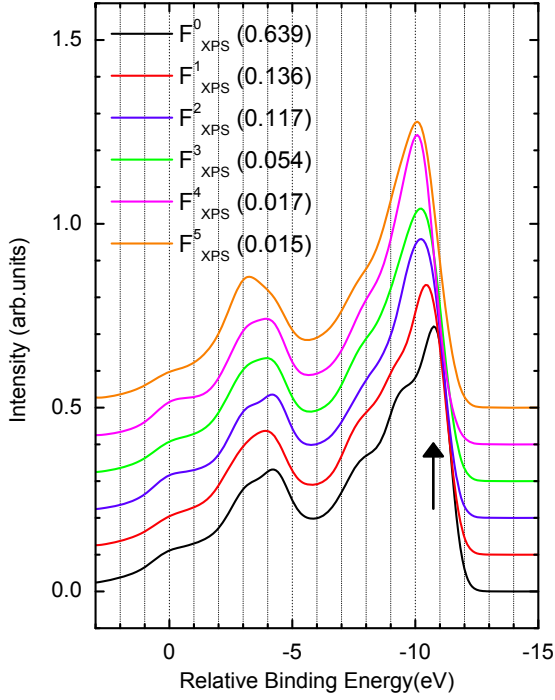}
\end{center}
\vspace{-0.5cm}
\caption{(Color online)
 The Co 2$p_{3/2}$ spectra in the AF phase before the thermal average.
 The corresponding Boltzmann factors at 300 K are shown in the parentheses.
 The spectral broadening is considered using a Gaussian of 0.5 eV width (HWHM).
}
\vspace{-0.4cm}
\label{fig:CoO_temp}
\end{figure}

Fig.~\ref{fig:CoO_temp} shows the contributions to the 2$p_{3/2}$ spectra
before the thermal average with the Boltzmann weights.
The $F^{0}$ in CoO shows single-peak ML with step-like high-$E_B$ tail.
The difference from NiO is mainly due to the richer multiplet structure
for Co$^{2+}$ than Ni$^{2+}$ \cite{okada92,groot04}.
Next, we observe a shift of the $F^n$ maxima towards higher $E_B$ in the spectra of the excited states.
This is attributed to a suppression of the NLS effect in the excited states 
by the same argument as in NiO.
The NLS induced shift together with thermal averaging is therefore instrumental for the formation
of the broad ML observed in the $2p$ XPS experiments.

\subsection{MnO}

In Fig.~\ref{fig:MnO}a we compare the calculated Mn 2$p$ XPS in PM MnO 
to the experimental spectra \cite{bagus00}.
The CT effect in 2$p$ XPS is known to be weaker in MnO compared
to NiO and CoO \cite{bagus00,okada92,taguchi97,bagus06}.
This fact is reproduced by the present result as well as previous CM calculations~\cite{okada92,taguchi97}.
To simulate the NLS effect we have performed a calculation with hybridization density
where the low-energy peak was artificially removed, see Fig.~\ref{fig:MnO}b. 
In the simulated spectrum, the CT satellite is almost identical to the full calculation,
while the low $E_B$ side of the ML is enhanced leading to a discrepancy with the experiment.
To our knowledge, the ML features in Mn$^{2+}$ have not been discussed in the context of NLS so far.
Very recently, Higashiya et al. \cite {higashiya17} performed
HAXPES measurements for LaOMnAs and (LaO)$_{0.7}$MnAs 
and found a sharp change in the Mn 2$p_{3/2}$ ML structure upon hole doping,
which calls for theoretical explanation.

\begin{figure}
\begin{center}
   \includegraphics[width=75mm]{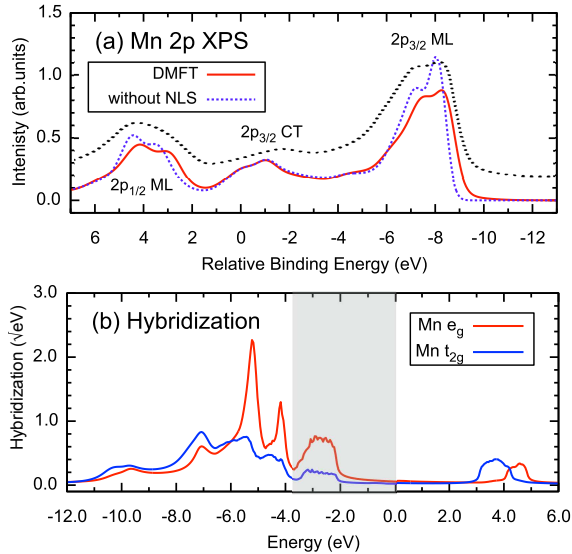}
\end{center}
\vspace{-0.5cm}
\caption{(Color online)
 (a) Mn 2$p$ XPS spectra in MnO calculated by the DMFT framework.
 A simulated spectrum ignoring the NLS effect from Mn 3$d$ band is shown together, for comparison.
 The spectral broadening is considered, using a Gaussian of 0.4 eV width (HWHM).
 The experimental data (dashed) is taken from Ref.~\cite{bagus00}.
 (b) The hybridization densities between the impurity Mn 3$d$ and the (host) valence states.
}
\vspace{-0.4cm}
\label{fig:MnO}
\end{figure}



\subsection{V$_2$O$_3$}
Unlike the Mott insulators studied so far, V$_2$O$_3$ is a paramagnetic metal under ambient conditions.
In Fig.~\ref{fig:V2O3}a we compare the V 2$p$ XPS with the experimental spectra of Ref.~\cite{taguchi05}.
The 2$p_{3/2}$ ML shows a characteristic broad-band structure with several shoulders.
To analyze the origin of the shoulders,
Figs.~\ref{fig:V2O3}b,c show the valence spectral densities and hybridization densities $V(\varepsilon)$, respectively.
In Fig.~\ref{fig:V2O3}b, the so-called lower Hubbard band, upper Hubbard band and the coherent peak at the 
chemical potential are obtained.
The coherent peak is characteristic for correlated metals \cite{keller04,georges96}.
Besides hybridization with the main O $2p$ band, $V(\varepsilon)$ in Fig.~\ref{fig:V2O3}c exhibits three small peaks
corresponding to the hybridization of V 3$d$ states on the impurity site with the Hubbard bands and coherent peak.
Although these features appear negligible compared to the O 2$p$ peak,
the charge screening from their part below $E_F$ is responsible for width and shape of the V 2$p_{3/2}$ ML.
Indeed, the shoulders disappears if the hybridization density above $-4.0$ eV is artificially removed
in the spectral calculation, as shown in Fig.~\ref{fig:V2O3}a.
Therefore, the V 2$p$ XPS is quite sensitive to the fine features near $E_F$.

\begin{figure}
\begin{center}
   \includegraphics[width=70mm]{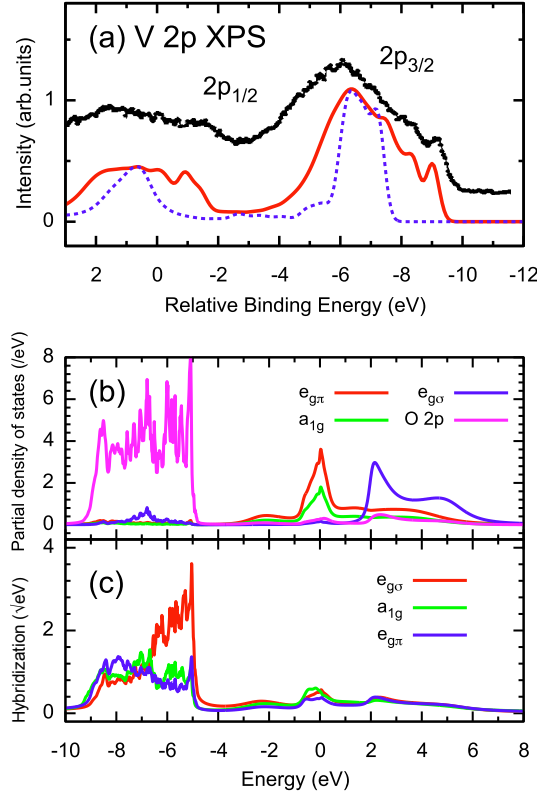}
\end{center}
\vspace{-0.5cm}
\caption{(Color online)
 (a) V 2$p$ XPS spectra in V$_2$O$_3$ calculated by the DMFT framework.
 The dashed line is a simulated spectrum ignoring the hybridization above $-4.0$ eV.
  The spectral broadening is considered using a Gaussian of 0.25 eV width (HWHM).
  The experimental data is taken from Ref.~\cite{taguchi05}.
 (b) Valence spectral intensities and 
 (c)  hybridization densities. 
}
\vspace{-0.4cm}
\label{fig:V2O3}
\end{figure}

\subsection{Fe$_2$O$_3$}

\begin{figure}
\begin{center}
   \includegraphics[width=75mm]{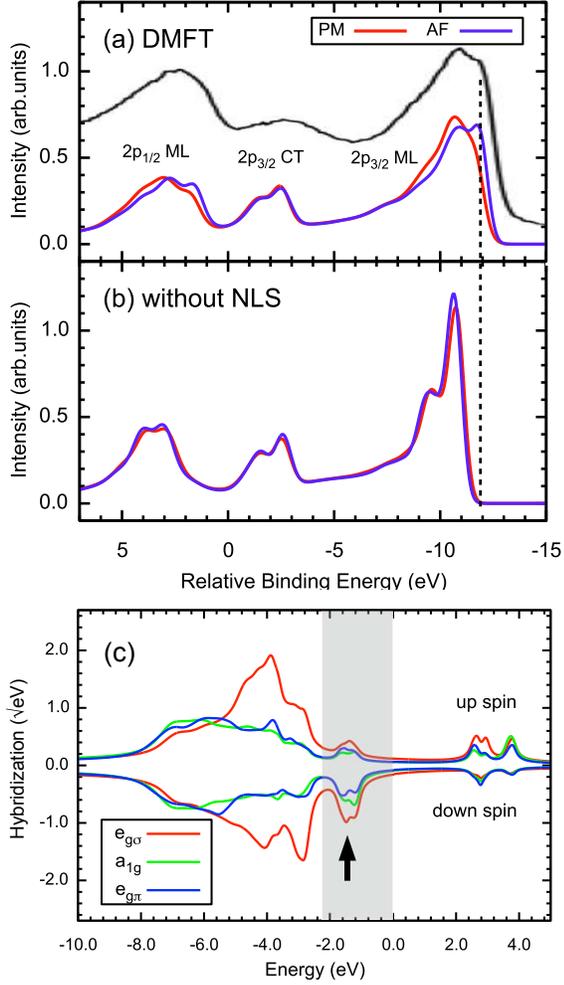}
\end{center}
\vspace{-0.5cm}
\caption{(Color online)
 (a) Fe 2$p$ XPS spectra in Fe$_2$O$_3$ for the AF and PM phase calculated by the DMFT framework.
   The experimental data is taken from Ref.~\cite{miedema15}.
 (b) Simulated spectra ignoring the NLS effect from Fe 3$d$ band.
 The spectral broadening is considered using a Gaussian of 0.4 eV width (HWHM).
 (c) The hybridization densities 
 between the impurity Fe 3$d$ and the (host) valence states for the up-spin unit in the AF phase.
}
\vspace{-0.4cm}
\label{fig:Fe2O3}
\end{figure}

Figure.~\ref{fig:Fe2O3}a shows  the Fe 2$p$ XPS in Fe$_2$O$_3$
in the AF and PM phase.
The overall structure of the spectra agrees well with experiments \cite{droubay01,fujii99,miedema15}. 
The interpretation of the Fe 2$p_{3/2}$ XPS in Fe$_2$O$_3$ so far has been controversial.
Droubay et al. \cite{droubay01} observed a double-peak feature in the Fe  2$p_{3/2}$ ML,
whereas Fujii et al. \cite{fujii99} observed a broad structure in the ML.
Though the ambiguity of ML might be caused by surface effects inherent to soft X-ray experiments.
Nevertheless, the double-peak like feature was observed also
in recent bulk-sensitive HAXPES experiments by Miedema et al \cite{miedema15},
but could not be their CM analysis.

\begin{figure}
\begin{center}
   \includegraphics[width=70mm]{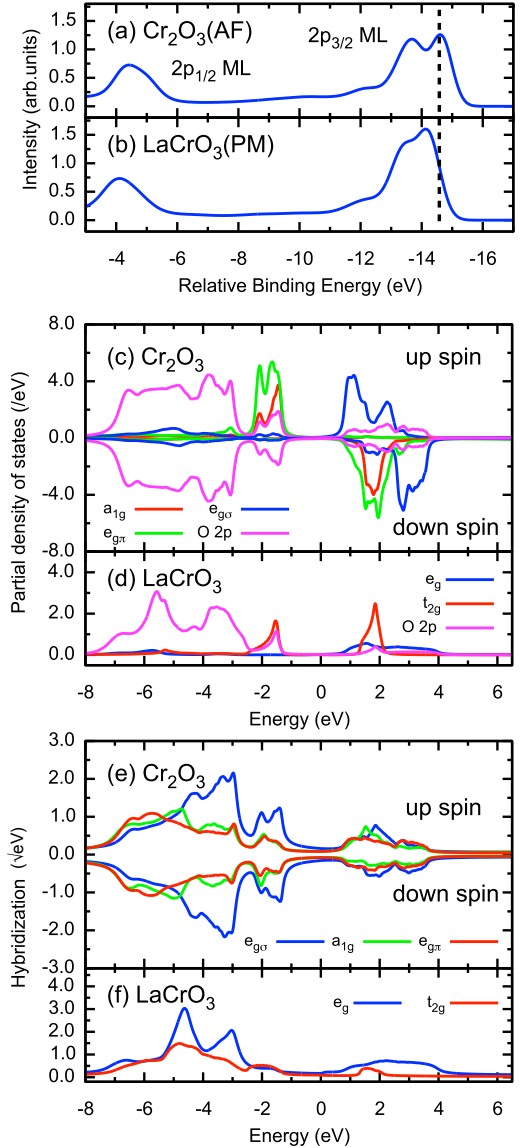}
\end{center}
\vspace{-0.5cm}
\caption{(Color online)
 Cr 2$p$ XPS spectrum
 in (a) Cr$_2$O$_3$ in AF phase and (b) LaCrO$_3$ in PM phase calculated by the DMFT framework.
  The spectral broadening is considered using a Gaussian of 0.4 eV width (HWHM).
 Valence spectral intensities calculated for (c) Cr$_2$O$_3$ and (d) LaCrO$_3$.
The hybridization densities
 between the impurity Cr 3$d$ state and the (host) valence states in (e) Cr$_2$O$_3$ and (f) LaCrO$_3$.
}
\vspace{-0.4cm}
\label{fig:Cr2O3}
\end{figure}


In Fig.~\ref{fig:Fe2O3}a, the double-peak structure of the 2$p_{3/2}$ ML is obtained in the AF phase,
which is in  a good agreement with the HAXPES data \cite{miedema15}.
We attribute the low $E_B$ part of the ML to the NLS from the Fe 3$d$ bands.
In Fig.~\ref{fig:Fe2O3}b we show the spectra without the NLS from Fe 3$d$ bands obtained
by artificially removing the shaded area of Fig.~\ref{fig:Fe2O3}c from the hybridization density.
Thus simulated Fe 2$p_{3/2}$ ML consists of a sharp peak with a high-$E_B$ shoulder  (multiplet effect), while the low-$E_B$ peak disappears
\footnote{The  2$p_{3/2}$ ML of the simulated spectra is similar the CM result of Ref. \cite{miedema15}.}.
Our result shows that the double-peak feature observed in the HAXPES experiments is an intrinsic feature of Fe$_2$O$_3$.
In the PM phase, the low $E_B$ peak is suppressed relative to the AF spectrum.
This suggests that the mechanism of polarization dependent NLS similar to NiO
is in effect also in Fe$_2$O$_3$.


\subsection{Cr$_2$O$_3$}

Finally, we discuss two Cr compounds with Cr$^{3+}$ valency
Cr$_2$O$_3$ and LaCrO$_3$, which are AF and PM insulators at room temperature, respectively.
In Fig.~\ref{fig:Cr2O3}a we compare their calculated  Cr 2$p$ XPS. 
The overall shape of the spectra is consistent with the soft X-ray experiments \cite{qiao13}.
Despite their almost identical calculated Cr $3d$ charge states and gaps
the Cr 2$p_{3/2}$ MLs have different shapes.

In order to explain the different 2$p_{3/2}$ MLs,
one needs to understand a relationship of the NLS to the crystal geometry. 
In both compounds the Cr $t_{2g}$ orbitals are half filled.
In LaCrO$_3$, the NLS originates in the occupied $t_{2g}$ states
on the neighboring Cr atoms, while the empty $e_g$ states cannot contribute.
\footnote{Note that local screening from the neighboring oxygen 2$p$ states to the $e_g$  states is possible.}
The NLS from the $t_{2g}$ states occurs via the $\pi\pi$-path,
which is quite weak in the perovskite structure. 
On the other hand, in Cr$_2$O$_3$ with the corundum structure, 
the  $t_{2g}$ states on neighboring Cr sites contribute the NLS to the $e_{g\sigma}$ states on the excited Cr site via a $\pi\sigma$-path
which is stronger than the $\pi\pi$-path in LaCrO$_3$.
This point is quantified in Figs.~\ref{fig:Cr2O3}e,f.  The hybridization densities at low
$E_B$ exhibit a pronounced difference with the LaCrO$_3$ one being substantially smaller than the Cr$_2$O$_3$ one.
As a result, the LaCrO$_3$ spectrum is relatively well described by the CM while in the Cr$_2$O$_3$ spectrum
NLS plays an important role. This demonstrates the sensitivity of the core-level XPS to the crystal geometry facilitated mainly by the NLS.
HAXPES measurements on Cr$_2$O$_3$ and LaCrO$_3$ as well as their doped versions are
highly desirable to test our findings.

\section{Conclusions and outlook}
\label{sec:5}

We have presented a systematic LDA+DMFT-based computational study of the 2$p$ core-level XPS in typical 3$d$ transition-metal oxides,
which was able to accurately reproduce the fine features observed in high-resolution experiments.
The non-local screening from 3$d$ states on the TM neighbors of the excited atom,
absent in the conventional analysis using the cluster model, was shown to be crucial for quantitative description 
of the studied spectra.
Our results show that the core-level XPS is sensitive to generically non-local effects such as lattice
geometry or magnetic order. However, to disentangle the non-local effects from the atomic multiplet effects in the XPS
theoretical simulations like the present one are necessary.
A combined theory\&experiment investigation of core-level XPS may provide insights into the physics of 
other classes of materials such as correlated metals, $4d$ and $5d$ materials with strong valence SO interaction,
or materials with more complicated geometries.

\begin{acknowledgments}
The authors thank F. de Groot, M. Ghiasi, J. Koloren\v{c}, M. Taguchi, M. Mizumaki, A. Sekiyama, H. Fujiwara, T. Saitoh, M. Okawa,
V. Pokorn\'y,  A. Sotnikov and J. Fern\'andez Afonso for fruitful discussions.
A. H thanks Y. Kawano, T. Yamamoto, Y. Ichinozuka, A. Yamanaka and K. Nakanishi for valuable discussions.  
A. H and J. K are supported by the European Research Council (ERC)
under the European Union's Horizon 2020 research and innovation programme (grant agreement No. 646807-EXMAG)
and T. U is supported by the JSPS KAKENHI Grant Number JP16K05407.
\end{acknowledgments}

\bibliography{xps_04_04}

\end{document}